# Molecule Generation and Optimization for Efficient Fragrance Creation

Bruno C. L. Rodrigues[a,b,c], Vinicius V. Santana[a], Sandris Murins[d], Idelfonso B. R. Nogueira[a,*]

[a]Chemical Engineering Department of the Norwegian University of Science and Technology, Gløshaugen, Trondheim, 7034, Norway
[b]LSRE-LCM–Laboratory of Separation and Reaction Engineering–Laboratory of Catalysis and Materials, Faculty of Engineering, University of Porto, Rua Dr. Roberto Frias, 4200-465 Porto, Portugal
[c]ALiCE–Associate Laboratory in Chemical Engineering, Faculty of Engineering, University of Porto, Rua Dr. Roberto Frias, 4200-465 Porto, Portugal
[b]SIA Murins Startups, Ausekļa iela 3 - 119, Riga, LV-1010, Latvia

## Abstract

This research introduces a Machine Learning-centric approach to replicate olfactory experiences, validated through experimental quantification of perfume perception. Key contributions encompass a hybrid model connecting perfume molecular structure to human olfactory perception. This model includes an AI-driven molecule generator (utilizing Graph and Generative Neural Networks), quantification and prediction of odor intensity, and refinery of optimal solvent and molecule combinations for desired fragrances. Additionally, a thermodynamic-based model establishes a link between olfactory perception and liquid-phase concentrations. The methodology employs Transfer Learning and selects the most suitable molecules based on vapor pressure and fragrance notes. Ultimately, a mathematical optimization problem is formulated to minimize discrepancies between new and target olfactory experiences. The methodology is validated by reproducing two distinct olfactory experiences using available experimental data.

## 1. Introduction

The human sense of smell, often overshadowed by its more celebrated counterparts like sight and hearing, is a quietly powerful faculty that profoundly shapes our lives. Beyond its role as a sensory pleasure, the olfactory sense is a sentinel of survival, a guardian against perils hidden in the air. It has guided us through the ages, allowing us to discern the delicious from the dangerous, the fragrant from the foul (Synnott, 1991).

This extraordinary sense has linked us to our ancient ancestors, whispering warnings of spoiled sustenance and signaling the presence of threats concealed in nature's bouquet. Specially after COVID-19, where many people lost their sense of smell and taste temporarily, and in some cases permanently (Mullol et al., 2020), it is evident how crucial this sense is, even if we may take it for granted.

Recent findings indicate that humans can distinguish more than a trillion unique scents (Bushdid et al., 2014). The intricacies of how we recognize and interpret these chemical compounds still hold some mystery. Nevertheless, our fascination with fragrances has given rise to a thriving industry over the centuries (Leffingwell & Associates, 2018). In 2017, the Flavors and Fragrances (F&F) industry was valued at over 26 billion US dollars, and despite an 8.4% market dip in 2020 due to the COVID-19 pandemic, projections estimate its worth to exceed 36 billion US dollars by 2029 (Fortune Business Insight, 2022). In 2023, the fragrance market shows no sign of slowing down, as most groups report double-digit growth and a significant rise in demand (Ludmir, 2023). This industry encompasses various products, ranging from everyday bar soaps to high-end perfumes, the latter being more exposed to consumer scrutiny, and being distinguished by higher value, and a sophisticated development process.

On the other hand, fragrances carry another important aspect for human society, it's heritage through time. Rituals, places, flowers, ecosystems are characterized by their specific fragrances, and their olfactory landscape composes an important element for cultural identity of a society. Documenting then is an important aspect usually neglected, due to many factors one of which is the lack of investment and studies addressing this aspect. The development of scents related technology, as is the case of this and previous works, can support the studies in olfactory heritage and promote the preservation of this important but neglected cultural asset (Bembibre & Strlič, 2017; Grau-Bové & Strlič, 2013; Herz et al., 2004).

A fragrance essentially comprises a blend of chemicals dissolved in one or more solvents, typically ethanol and water, which creates the distinct fragrance notes we perceive. These notes are generally described using words, or descriptors, and are categorized into three levels: top, middle, and base (Carles, 1961). Top notes, though captivating, are short-lived, providing the initial impression of the perfume and fading within minutes. The heart or middle notes form the core scent, lingering for several hours. Base notes, lasting from hours to days, can be likened to the aftertaste in food or drink. The persistence and strength of each fragrance note depend on factors like thermodynamics, transport phenomena, and psychophysics (A. E. Rodrigues et al., 2021). For a scent to be detected, molecules must become airborne and reach our nasal receptors. Numerous studies delve into psychophysical models that can evaluate human responses to evolving scents (Almeida et al., 2019; Mata et al., 2005; M. A. Teixeira et al., 2009; Wakayama et al., 2019). Within this framework, the emerging field of Perfume Engineering has taken root (A. E. Rodrigues et al., 2021).

Despite the consistent growth of the perfume industry, the importance of olfactory heritage and the emergence of Perfume Engineering, there is a noticeable gap in scientific literature providing innovative tools for this field. The predominant approach for creating or reproducing fragrances heavily relies on the traditional trial-and-error method blended with expert knowledge, this approach is time-consuming and expensive (Santana et al., 2021).

Currently, the F&F industry offers approximately 10,000 essential oils and scents. Developing a new perfume typically involves conducting around 1,000 tests using these available fragrances. This intricate process can span up to three years and incur costs of up to 50,000 US dollars per kilogram of the perfume (A. E. Rodrigues et al., 2021).

Nevertheless, the field has seen advancements. Many studies are emerging addressing these issues from their chemical aspects (Bembibre & Strlič, 2017, 2022; Curran et al., 2018; Prosen et al., 2007). Others are addressing it with a point of view of chemical engineering (Almeida et al., 2019; Mata et al., 2005; A. E. Rodrigues et al., 2021; Rodríguez et al., 2011; M. Teixeira et al., 2013; M. A. Teixeira et al., 2009, 2014). The new trend is to bridge these aspects in a concise framework making use of AI and advanced computing science. Several studies have explored the potential of using machine learning in combination with natural language processing to classify the scent of unidentified molecules (Debnath & Nakamoto, 2022; Gerkin, 2021; Nozaki & Nakamoto, 2018; Saini & Ramanathan, 2022). While this approach expands the palette of fragrances for creators, it doesn't offer insights into optimal fragrance combinations. Santana et al., 2021 have advocated machine learning as an alternative to phenomenological models for perfume engineering. Their work opens the door to exploring numerous combinations for fragrance creation. The present work builds on this knowledge and attempts to create a novel tool that may help perfume creators in their craft.

Research from other fields reveals that computer-aided molecular design can predict how a molecule relates to a desired property (Heng et al., 2022; Ooi et al., 2022; Queiroz et al., 2023c, 2023b, 2023a; Radhakrishnapany et al., 2020; Sanchez-Lengeling et al., 2019; Sharma et al., 2021; Zhang et al., 2018). This insight opens new possibilities for the perfume industry, such as the potential to engineer molecules specifically tailored to emit desired fragrances. In our previous work, a methodology was developed to generate fragrance molecules that fit into a desired target smell. Implementing this approach streamlines the vast array of molecular options down to a select few optimal ones, reducing the time of development of a new fragrance substantially (B. C. L. Rodrigues et al., 2023). However, the composition is not determined. Hence, it would still need to be tested and optimized through trial and error. In this work, a novel approach is adopted to generate molecules that fit a desired target smell, going one step further, and indicates what is the ideal composition of solvents and these newly generated molecules, resulting in a complete recipe for perfume production.

In this context, the main contributions of this work can be listed as follows:

- Proposing a quantitative analysis of the link between perfume molecular structure and human olfactory perception, facilitating scent psychophysics knowledge.
- Development and validation of an AI-driven molecule generator customized for predicting scent molecules, supporting fragrance chemistry field.
- Proposing of an optimization problem to reproduce fragrance perception by finding substitute components and their compositions.
- Fine-tuning the optimal combination of solvents and newly generated molecules to achieve the desired fragrance.
- Potential reduction of the time-consuming and expensive trial-and-error method commonly associated with perfume production.

Overall, this work proposes a framework based on AI and phenomenological knowledge to reproduce olfactory experience given a target fragrance. The work then finds molecular substitutes and proposes a new formulation.

## 2. Methodology

This study is based on an approach that combines data analysis and computational methodologies to develop the field of perfume engineering and fragrances reproduction. Our framework is comprehensive, with the goal of maximizing the potential of available data and making computer assisted judgments in the creation, reproduction and understanding of fragrances. The steps involved in our process are four-fold and are outlined in detail as:

I. Quantification of Target Scent: To begin, we perform a precise quantification of the desired scent targets. This involves taking known target fragrances and quantifying their scent by first calculating the equilibrium concentrations of fragrance molecules in the gaseous phase, then calculating the odor values for all constituent molecules, a concept that describes the intensity of a given smell. We then extend this process to compute the odor values of distinct scent families within the composition. An important issue in this step is to find a common vocabulary to describe the scent descriptors that composes the targeted fragrance. This is not an easy task and was not the focus of this work. Hence, we considered a vocabulary used in a previous work that describes the target (M. A. Teixeira et al., 2014).

II. Molecule Generation for Desired Perfume Profile: The core of our approach revolves around the goal of producing a collection of molecules that reproduces a given olfactory experience. This task unfolded in two key phases:

   a. Training Phase: We initiated the process by training a gated graph neural network (GGNN) using a carefully curated database encompassing well-documented molecules and their associated fragrance notes within a known descriptors vocabulary. This is similar to what has previously been done in other works (Sanchez-Lengeling et al., 2019).
   b. Generation and Enhancement: Employing the knowledge acquired through GGNN training, we generate a diverse array of molecules. To enhance and refine this generation process, we harnessed transfer learning techniques, focusing on molecules that align with our predefined scent profiles. A similar methodology was described in previous published works (Queiroz et al., 2023b, 2023c).

III. Selection of Best Fitting Molecules: Next, we focus on selecting molecules that align with the target fragrance profile. This step involves estimating the vapor pressure of the generated molecules and strategically choosing those with vapor pressures that closely match the original molecules. In this selection process, the fragrance notes of the chosen molecules are also considered, ensuring that the desired fragrance profile is effectively achieved.

IV. Minimizing mismatch between new and target scents: The final stage centers on optimizing the composition to harmonize with the defined scent targets. The aim here is to minimize the disparity between the calculated family odor values and the target odor values, thereby fine-tuning the fragrance composition to a level of utmost precision and alignment with the desired scent profile.

Steps III and IV of the methodology are the most significant contributions of this work. They build on the contributions of previous works and allow for a new application in the context of scent reproduction. The nuances of this proposed methodology and its implications will be delved into in the subsequent sections of this work. Figure 1 presents a schematical representation of the methodology proposed here and afore described.

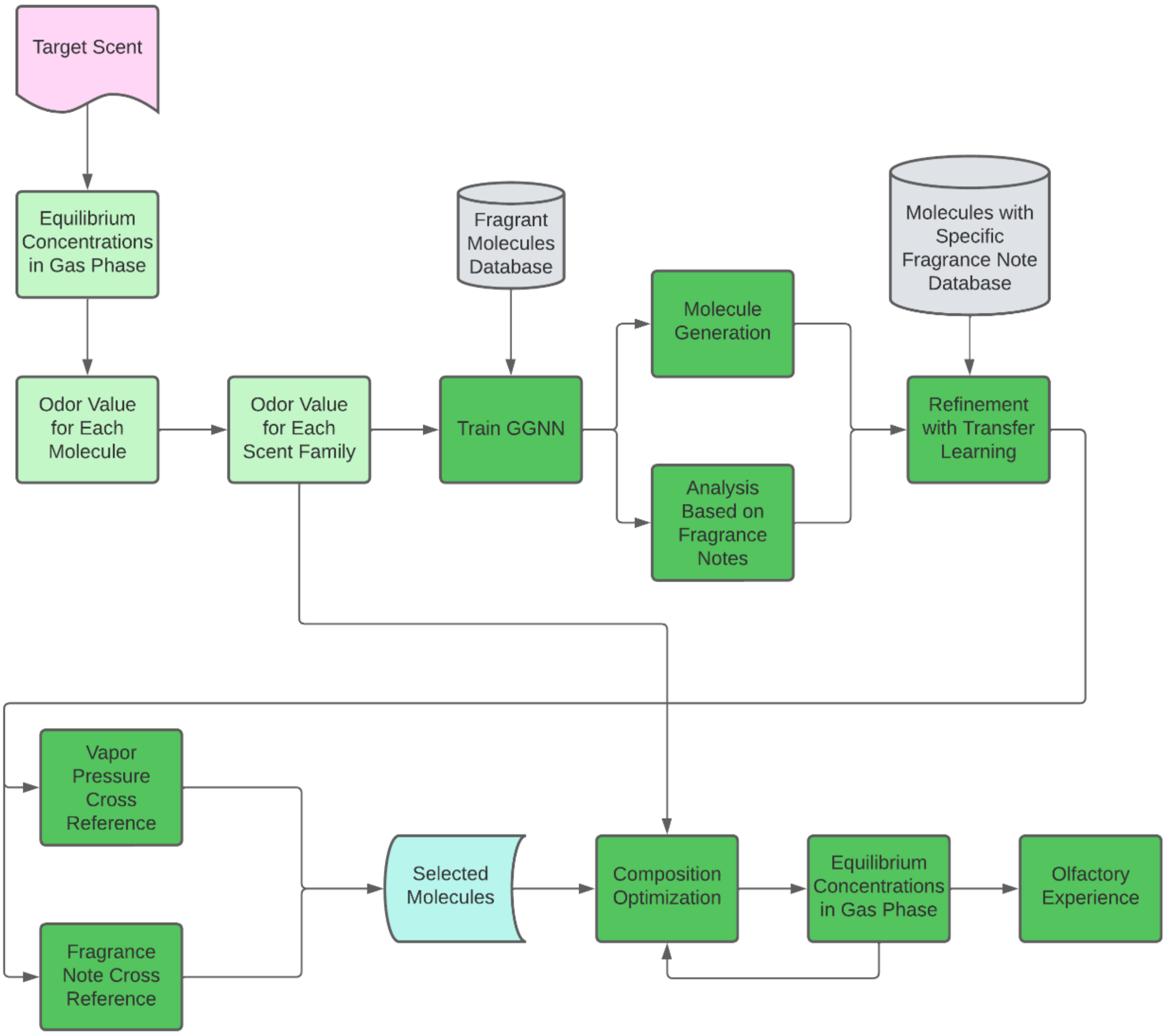

Figure 1 – Schematic representation of the methodology, including all four stages.

## 2.1. Quantification of Target Scent

Before any molecule can be generated and any fragrance composition can be optimized, it is necessary to define a target smell to be achieved. This target could be any fragrant mixture given the molecular composition. To transform such a subjective perception that is the smell of a mixture into numbers, the concept of odor values (OV) is employed (Poucher, 1955; M. A. Teixeira et al., 2014).

An odor value gives the intensity of a particular smell by calculating the ratio between the concentration of a fragrant molecule in the gas phase ($C_{g_i}$) and the minimum concentration necessary for the smell to be detected by the human nose, known as the odor detection threshold (ODT), which is defined as the minimum concentration where the molecule can be detected by the nose (Appell, 1969).

$$OV = \frac{C_{g_i}}{ODT} \tag{1}$$

The gaseous phase above the liquid (called the headspace, where the nose will detect the smell) can be treated as an ideal phase because gas molecules are highly diluted. However, in the liquid phase, non-idealities must be taken into consideration due to the presence of molecular interactions. Given these assumptions, the connection between both phases can be described through the lens of the modified Raoult's Law (M. A. Teixeira et al., 2010). The concentration of the odorant species in the headspace can be translated as follows:

$$C_{g_i} = \frac{y_i \cdot P \cdot M_i}{RT} = \frac{x_i \cdot \gamma_i \cdot P_i^{sat} \cdot M_i}{RT} \tag{2}$$

To quantify the olfactory experience of a perfume, an odor value is calculated for each of the fragrant molecules present in the perfumes. Then, the odor values of molecules ($OV_i$) with the same fragrance notes are combined to form the odor values of scent families ($OV_j$). Here, the fragrance notes are also divided into two levels: primary and secondary notes, as suggested by (M. A. Teixeira et al., 2010). The primary notes are the most identifiable notes in a smell and contribute more to the $OV_J$. Secondary notes represent the nuance of the smell and contribute less to the $OV_J$. A weight of 70% for primary notes and 30% for secondary notes is suggested in M. A. Teixeira et al., 2010 work, and is also considered here. Equations (3) and (4) describe how these values are calculated.

$$OV_i = \gamma_i x_i \left(\frac{P_i^{sat} M_i}{ODT_i}\right) \cdot \left(\frac{1}{\mathrm{RT}}\right) \tag{3}$$

$$OV_j = \frac{\sum OV_{i,primary} \cdot 0.7 + \sum OV_{i,secondary} \cdot 0.3}{\sum OV_i} \tag{4}$$

In addition to the molecular composition of the mixture, it is necessary to know readily available information such as the molecules' vapor pressure and molecular weight, as well as the ambient temperature. Furthermore, the activity coefficient of each molecule must be calculated; here the UNIFAC method was used (Poling et al., 2004). The functional groups of the molecules are identified so that the molecule can be segmented into groups. Then, the activity coefficient is calculated by estimating the interaction of groups. For this work, the method was executed in the Julia programming language, using the Clapeyron.jl and the GCIdentifier.jl libraries(Walker et al., 2022).

The last remaining information that is necessary is the odor detection threshold (ODT). Values for ODT can be difficult to obtain, however, there are published compilations of ODT for several molecules available in the literature (Devos et al., 1990; van Gemert, 2003). However, since these compilations are not guaranteed to contain the molecules that are generated in this work, an estimation method, as described by Rodríguez et al. in 2011, was also and will be explained in further detail in section 2.4. of this document.

Two fragrances with known compositions and all the necessary information were obtained from a published work (M. A. Teixeira et al., 2014) and used as a case study for this research. Since the fragrance notes of each molecule were not available in the source used, The Good Scents Company (2021) was chosen as a reference to consult the fragrance profile of all compounds. Primary fragrance notes, which represents the most identifiable note from the compound and secondary fragrance notes, representing the nuance of the smell were retrieved from the good scents company. The data is detailed in Table 1.

Table 1 – Composition of the target fragrance in liquid phase and additional molecular information

| Molecule | MW (g/mol) | $P_{sat}$ (Pa) | ODT ($mg/m^3$) | Primary Fragrance Note | Secondary Fragrance Note | Molar Fraction of First Fragrance | Molar Fraction of Second Fragrance |
|---|---|---|---|---|---|---|---|
| (+)-(R)-limonene | 136.2 | 205 | $2.45 \times 10^{0}$ | Citrus | Fruity | $1.41 \times 10^{-2}$ | $5.20 \times 10^{-4}$ |
| Benzyl Acetate | 150.2 | 189 | $3.32 \times 10^{-1}$ | Sweet | Floral | $3.56 \times 10^{-4}$ | $4.89 \times 10^{-4}$ |
| Eugenol | 164.2 | 3.01 | $2.56 \times 10^{-3}$ | Sweet | Spicy | $2.68 \times 10^{-3}$ | $9.23 \times 10^{-3}$ |
| Coumarin | 146.1 | 0.13 | $3.09 \times 10^{-4}$ | Coumarinic | Sweet | $2.27 \times 10^{-3}$ | $1.85 \times 10^{-3}$ |
| Citral | 152.2 | 9.49 | $3.69 \times 10^{-2}$ | Sharp | Lemon | $7.38 \times 10^{-3}$ | $4.29 \times 10^{-4}$ |
| Geranyl Acetate | 196.3 | 3.47 | $1.47 \times 10^{1}$ | Floral | Rose | $6.11 \times 10^{-3}$ | $8.49 \times 10^{-3}$ |
| Styrallyl Acetate | 164.2 | 27.1 | $5.01 \times 10^{-2}$ | Green | Fruity | $7.49 \times 10^{-3}$ | $5.76 \times 10^{-4}$ |
| Musk Ketone | 294.3 | 0.0016 | $3.46 \times 10^{-4}$ | Fatty | Musk | $2.59 \times 10^{-3}$ | $9.55 \times 10^{-4}$ |
| β-ionone | 192.3 | 2.27 | $2.09 \times 10^{-2}$ | Floral | Woody | $6.22 \times 10^{-3}$ | $4.17 \times 10^{-3}$ |
| Water | 18.0 | 3170 | Odorless | Odorless | Odorless | $2.80 \times 10^{-1}$ | $3.99 \times 10^{-1}$ |
| Ethanol | 46.0 | 7270 | $5.53 \times 10^{1}$ | Alcoholic | Alcoholic | $6.71 \times 10^{-1}$ | $5.74 \times 10^{-1}$ |

## 2.2. Molecule Generation for Desired Perfume Profile

### a. Training Phase

Having obtained the target smell it is now necessary to obtain potential molecules that can reproduce the target. In this context, Gated Graph Neural Networks (GGNNs), tailored for molecular synthesis, offer a powerful approach. By training a GGNN on an established database of fragrant molecules, this network gains the capability to generate novel molecules while drawing inspiration from the characteristics of existing compounds.

To train the GGNN effectively, access to a database containing known molecules with their associated fragrance notes is vital. This database allows the model to learn which molecular features are linked to specific fragrance notes. The database is sourced from The Good Scents Company (2021). The database includes the SMILES (Weininger et al., 1989) representation of molecules, which are then used by the model. The database used for this work can be found on the following GitHub repository: github.com/BrunoCLR/MolGen_Opt.

The GGNN requires graphs as inputs to generate new graphs on a later phase. For this effect, the transformation of SMILES representations into graph structures is carried out using functions from the RDKit library, an open-source cheminformatics software (*RDKit: Open-Source Cheminformatics*, n.d.). This transformation interprets atoms as nodes and chemical bonds as edges, embedding crucial chemical information into the graph. It also encompasses features such as the adjacency matrix, illustrating node connections, and edge attributes denoting distances between graph edges.

Molecules, now represented as graphs, undergo preprocessing to enable the generative network to reconstruct them effectively during the generation process. Each node and edge are numerically labeled based on its type. A starting node is determined, and molecules are sequentially deconstructed in accordance with the label order. These processed molecules then serve as inputs for GGNN training.

The GGNN, an advanced version of the graph neural network introduced by (Li et al., 2015) incorporates gated recurrent units (GRU) in its propagation phase and utilizes back-propagation through time (BPTT) as introduced by (Zhou et al., 2019). This integration of GRU and BPTT tackles the vanishing/exploding gradient challenge, effectively addressing diminishing gradients as the network delves deeper into the recurrent neural network. The GRU employs update and reset gates to discern and prioritize pertinent data for predictions, enhancing the model's predictive capabilities based on historical data. Simultaneously, BPTT optimizes performance chronologically, adapting traditional backpropagation methods for systems with memory, exemplified by the GGNN.

The GGNN produces two primary outputs: the graph embedding (g) and the final transformed node feature matrix (HL). These outputs subsequently serve as inputs for the global readout block (GRB), structured as a multi-layered perceptron (MLP) architecture. The GRB calculates the action probability distribution (APD) for the graph, comprising probabilities for various actions to evolve the graph. These action probability distributions guide the model in the process of graph creation, with potential actions including introducing a new node to the graph, linking the most recent node to a pre-existing one, or completing the graph.

It is essential to note that certain actions may be unsuitable for specific graphs, necessitating the model's ability to assign zero probabilities to such invalid actions. The cumulative probabilities of all actions must sum to 1, establishing the target vectors that the model seeks to learn during training.

In essence, the GGNN, supported by a robust database of fragrant molecules, transforms SMILES representations into graph structures, using the described neural network techniques to generate novel molecules inspired by known compounds. This approach holds significant promise for the field of fragrance chemistry, offering a data-driven method for designing new fragrances and exploring the molecular intricacies of scents.

$$h^{0}{}_{v} = x_{v} \tag{5}$$

$$r_v^t = \sigma\left(c_v^r \sum\nolimits_{u \in N_v} W_{le}^r h_u^{(t-1)} + b_{le}^r\right) \tag{6}$$

$$z_v^t = \sigma\left(c_v^z \sum\nolimits_{u \in N_v} W_{le}^z h_u^{(t-1)} + b_{le}^z\right) \tag{7}$$

$$\tilde{\mathrm{h}}_{\mathrm{v}}^{\mathrm{t}} = \rho(\mathrm{c}_{\mathrm{v}} \sum\nolimits_{\mathrm{u} \in \mathrm{N}_{\mathrm{v}}} \mathrm{W}_{\mathrm{le}} \left(r_u^t \odot h_u^{(t-1)}\right) + b_{le} \tag{8}$$

$$h_v^t = (1 - z_v^t) \odot h_v^{(t-1)} + z_v^t \odot \tilde{\mathrm{h}}_{\mathrm{v}}^{\mathrm{t}} \tag{9}$$

Where $h^{0}{}_{v}$ is the node feature vector for the initial node $v$ at the GGNN layer and is equal to its node feature vector in the graph, $r_v^t$ is a GRU gate in the specific MLP layer $t$, and relative to the node $v$, $c_v = c_v^z = c_v^r = |N_v|^{-1}$are normalization constants, $N_v$ is the set of neighbour nodes for $v$; u is a specific node in the graph, $W_{le}^r$ is a trainable weight tensor in $r$ regarding the edge label $le$, $b$ is a learnable parameter, $z$ is also a GRU gate, $\rho$ is a non-linear function, $\odot$ is an element-wise multiplication. The functional form of these equations is translated by the following:

$$h^{0}{}_{i} = x_{i} \tag{10}$$

$$m_i^{l+1} = \sum\nolimits_{vj \in N_{(vi)}} MLP^e\left(h_j^l\right) e_{ij} \tag{11}$$

$$h_i^{l+1} = GRU\left(m_i^l + 1\,, h_i^l\right) \tag{12}$$

$$\forall l \in L \tag{13}$$

Where $m_i^{l+1}$ and $h_i^{l+1}$ are the incoming messages and hidden states of node $vi$, $eij$ is the edge feature vector between $vi$ and $vj$, $l$ is a GNN layer index and $L$ is the final GNN layer index. $g,$ the final graph embedding is given by:

$$g = \sum\nolimits_{vi \in v} \sigma(MLP^a(h_i^L)) \odot \tanh\,(MLP^b([h_i^L, h_i^0]) \tag{14}$$

The processes undertaken by the global readout block are translated by the following equations. The SOFTMAX function is the activation function of the block, it converts a vector of numbers into a vector of probabilities.

$$f'_{add} = MLP^{add,1}(H^L) \tag{15}$$

$$f'_{conn} = MLP^{conn,1}(H^L) \tag{16}$$

$$f'_{add} = MLP^{add,2}([f'_{add}, g]) \tag{17}$$

$$f_{conn} = MLP^{conn,2}([f'_{conn}, g]) \tag{18}$$

$$f_{add} = MLP^{fin,2}(g) \tag{19}$$

$$APD = SOFTMAX([f_{add}, f_{conn}, f_{fin}]) \tag{20}$$

The training phase is performed in small batches, employing the scaled exponential linear unit (SELU) as the activation function, which is applied after each linear layer in the MLP. The loss function during this phase is calculated as the Kullback-Leibler divergence (Kullback & Leibler, 1951) between the target and predicted action probability distribution (APD). Furthermore, the model utilizes the Adam optimizer at various stages, a well-established first-order gradient-based optimization algorithm introduced by (Kingma & Ba, 2014).

Throughout the training phase, regular graph samples are taken for evaluation at consistent intervals. The chosen metric for this assessment is the uniformity-completeness Jensen-Shannon divergence (UC-JSD), as introduced by Arús-Pous et al. in 2019. UC-JSD acts as a gauge of the similarity between the distributions of the negative log-likelihood (NLL) per sampled action. Ideally, these values should tend towards zero.

The last phase is dedicated to graph generation. In this stage, the APD, formed in the global readout block (GRB), is utilized to construct graphs. A graph continues to grow until either the 'terminate' action is selected from the APD or an invalid action occurs. Invalid actions encompass adding a node to a non-existent node in a graph (unless the graph is vacant), connecting an already linked pair, or appending a node to a graph that has reached its node capacity, as determined during preprocessing. It's essential to highlight that hydrogens are excluded in both the training and generation stages, with subsequent incorporation using RDKit functions based on the valency of each atom.

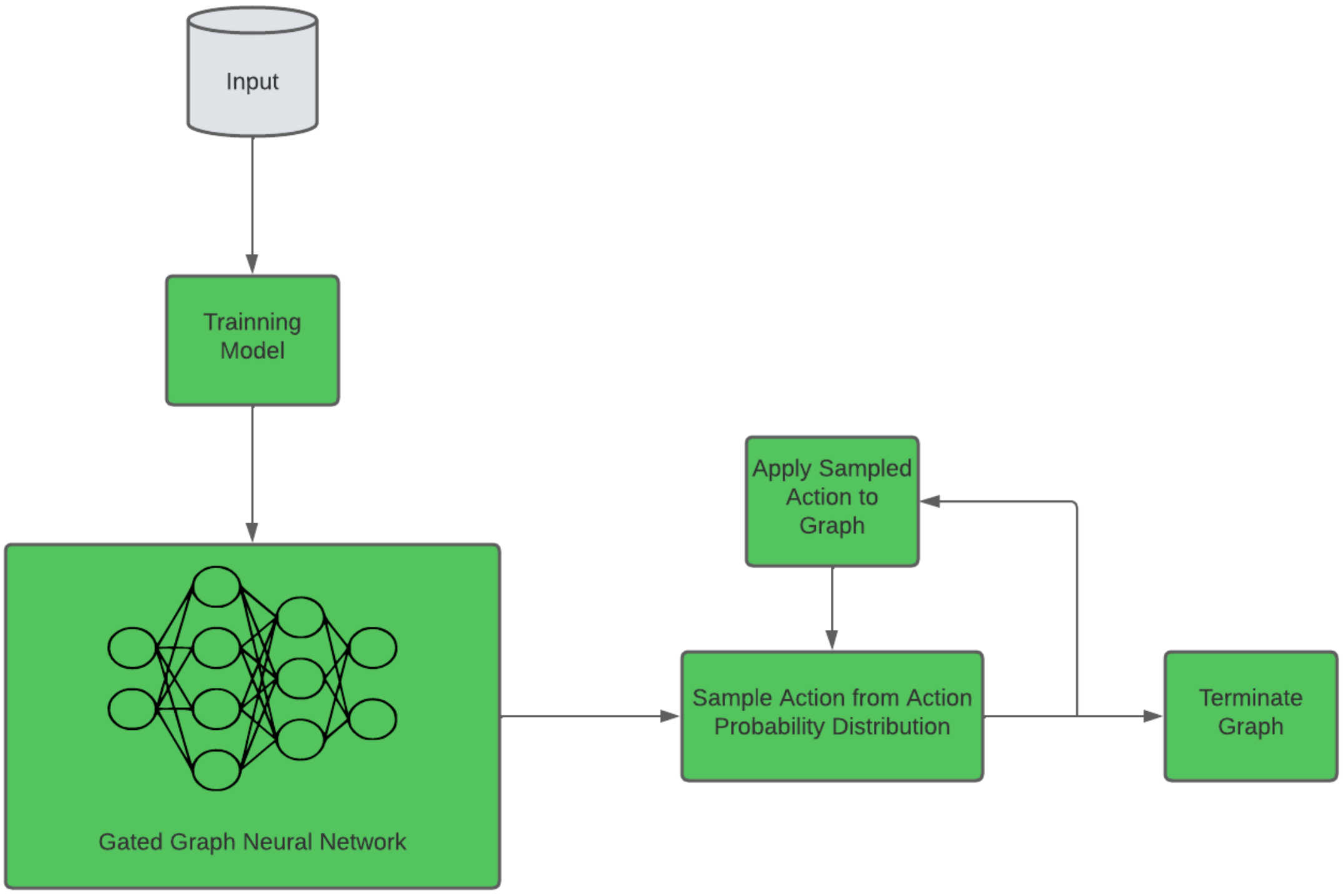

Figure 2 – Schematic representation of the methodology of the GGNN model.

## b. Generation and Enhancement

The previously generated molecules were all likely to be fragrant, given that the GGNN model was exclusively trained on a database comprising fragrant molecules. However, the model's output lacks information about the specific fragrance notes associated with each molecule. As the primary goal of this endeavor is to identify ingredients suitable for crafting perfumes, the generation of molecules corresponding to the desired fragrance notes becomes imperative. To achieve this, a technique known as transfer learning was employed.

The concept of transfer learning, as defined by Xue et al. in 2019, entails a machine learning approach used to address problems with limited data by leveraging knowledge from another related problem or dataset. In this scenario, the insights acquired by the GGNN model during its comprehensive training on the entire database were transferred to a new model. This model, however, underwent training using molecules known to possess the desired fragrance notes.

Training the model exclusively with the known molecules from the outset was infeasible due to the insufficient number of molecules associated with each fragrance note in the database, which would compromise the generation accuracy. Instead, the previously trained model was retrained iteratively with a fresh set of molecules. This iterative process was repeated to generate molecules corresponding to all the target scents. The databases with specific fragrance notes can be found here: github.com/BrunoCLR/MolGen_Opt.

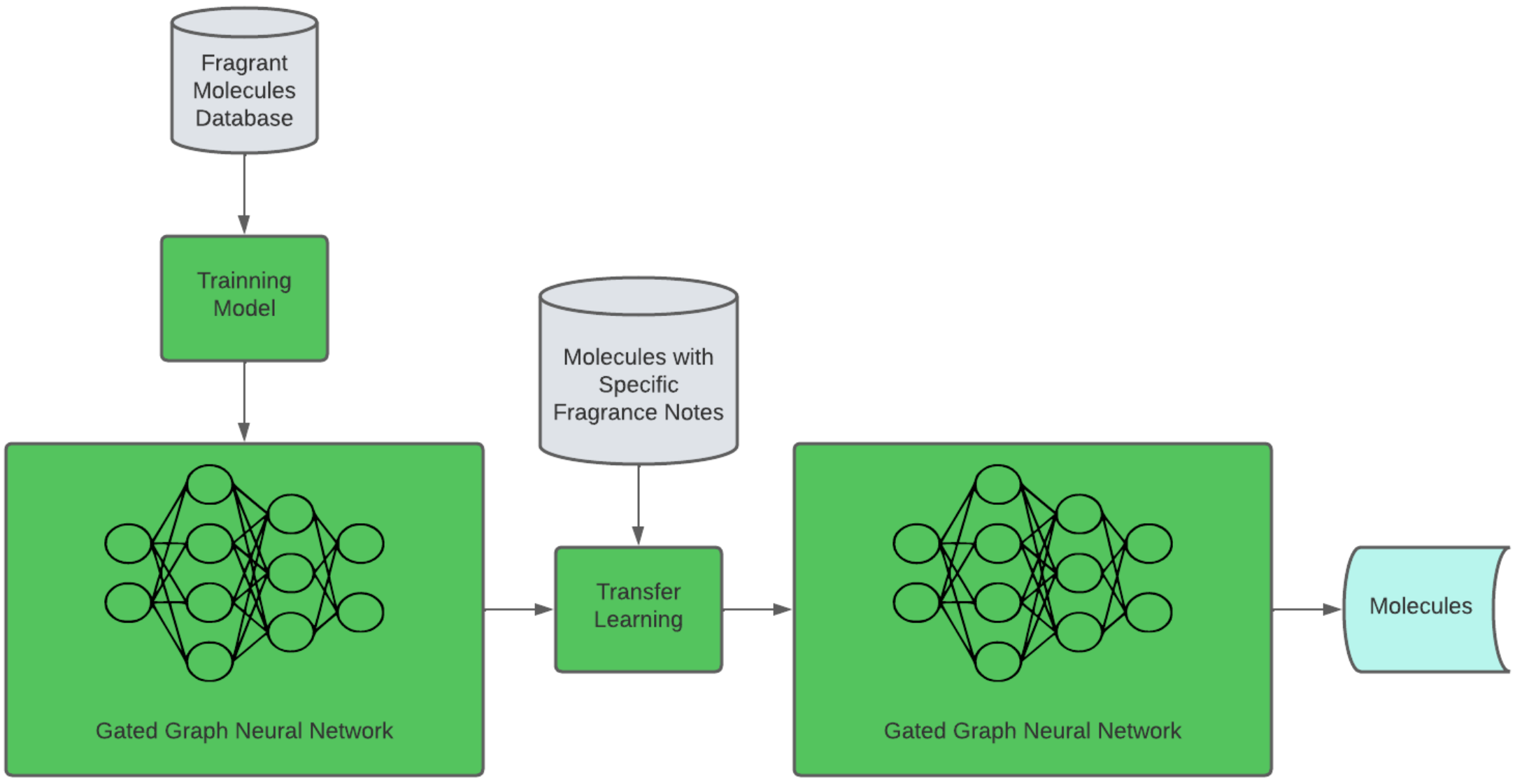

Figure 3 - Schematic representation of the transfer learning process.

## 2.3. Selection of Best Fitting Molecules

In the proposed process of generating molecules to match specific fragrance profiles, a given olfactory experience, several crucial steps are taken to ensure the quality and suitability of the compounds. The initial phase involves filtering the generated molecules. To do this, we employ predictive techniques to estimate the vapor pressure of each generated molecule. The thermo library, available in the Python programming language, is used for this effect (Bell, 2023). The library works by consulting databases of thermodynamic databases and using the parameters to calculate $P_{sat}$ using well-established thermodynamic correlations such as the Lee-Kesler, Ambrose-Walton, and other available correlations (Ambrose & Walton, 1989; Lee & Kesler, 1975; Poling et al., 2004). This method is quick and provides accurate estimations, however, it depends on the availability of thermodynamic parameters for each molecule being analyzed. For that reason, all generated molecules that do not have known thermodynamic coefficients are excluded from further analysis.

Vapor pressure is an essential parameter as it influences how readily a substance evaporates, contributing to its aroma and affecting the odor value. Molecules with vapor pressure values that closely resemble those of the original molecules are considered favorable candidates. While several molecules may give off the same aroma, they will last for different amounts of time, and the vapor pressure of each molecule affects that. By using the vapor pressure of the reference molecules as targets for the new molecules, the options are reduced to more suitable molecules. Furthermore, we conduct an additional check by referring to the fragrance notes of the generated molecules within the extensive database provided by (*The Good Scents Company*, 2021). This step ensures that the aroma of the selected molecules aligns with our desired fragrance profile. It serves as an additional layer of verification, making certain that the olfactory aspects match our expectations. Enough molecules are selected to ensure that all scent families are represented by the new mixture.

With a curated collection of molecules that meet our criteria, we proceed to the next significant step: the calculation of odor detection thresholds (ODT). Determining ODT values can be a challenging task, particularly for molecules that may not be widely studied or documented in the literature. Despite our best efforts to gather data on the ODT for each generated molecule, we encountered gaps where information was simply unavailable. To address this, we turned to an estimation method introduced by (Rodríguez et al., 2011). Their method, which accounted for 77% of the variance in odor thresholds of a test set of molecules, leverages readily available information, including the vapor pressure of each molecule, its solubility in water ($C_{i,W}$), the ambient temperature, and the octanol-water partition coefficient ($K_{OW}$). While a more precise estimation method might yield superior results, given the limited availability of data in the literature, this method stands as an adequate and practical solution. Our source of data for $C_{i,W}$ and $K_{OW}$ values is the extensive *ChemSpider* database by the (Royal Society of Chemistry, n.d.), ensuring that we have the necessary information for all the selected molecules in our study. Equation (21) details the calculation of the ODT values, allowing us to proceed to the optimization problem of fine-tuning our fragrance formulation.

$$\log(ODT)_i = 0.97 \cdot \log\left(\frac{P_i^{sat}}{K_{OW} C_{i,W} RT}\right) + 4.2 \tag{21}$$

## 2.4. Minimizing mismatch between new and target scents

Having obtained the necessary molecules to emulate the desired smell is already a big step towards fragrance formulation and reproduction, however, it is not enough. A pivotal aspect of the formulation process involves calculating the composition of the fragrance mixture to ensure it resonates harmoniously with the intended fragrance profile.

This work proposes an optimization problem that is driven by the aspiration of minimizing any disparities between the calculated family odor values and the predefined target odor values, the olfactory experience to be achieved. However, the task is far from straightforward, primarily due to the complexity and non-linearity of the underlying optimization problem and the system.

At the core of fragrance composition optimization lies the imperative to minimize a defined objective function. In this work the objective function was designed to minimize the mean squared error between two essential components: the target odor values (OV) and the calculated odor values of the potential scent families, as described earlier. The target values, represented as a vector consisting of 14 distinct elements, representing the odor values of the scent families of the original fragrance. The calculated values are the odor values corresponding to the current mixture. These values are calculated iteratively and vary in a non-linear manner with respect to the composition of the current mixture. The optimization problem is summarized by equations (22) to (26).

The molar fractions of the new mixture are the decision variables that the algorithm iteratively changes to minimize the objective function. Soft constraints regarding the final composition of the new mixture are included as penalties to the objective function. Equation (25) states all the included penalties: the first ensures the sum of all molar fractions add to one, averting invalid mixtures. The second and third penalties were included to avoid over-representation of two of the scent families, averting unwarranted deviations from the target that may arise in their absence. The scent families chosen for the penalties should be the ones with the highest odor values, or the ones that present the largest deviation from the target value. Finally, the last penalty constrains the OV of the alcoholic family to remain below a predefined threshold, maintaining consistency and conformity to established standards.

Moreover, the composition of the optimized fragrance must also respect box constraints, as stated in equation (26). These constraints, designed to safeguard the representation of each molecule within the mixture, prevent any potential misrepresentation or over-representation. A minimum value of $1\text{x}10^{-6}$ and a maximum value of 0.3 were considered for the fragrant molecules. By adhering to these specified box constraints, the composition strikes an equilibrium where each constituent molecule contributes its distinctive aroma without disproportionate influence.

$$\min_{\mathrm{i}} F_{obj}(x_i) \tag{22}$$

$$F_{obj} = MSE + Penalty \tag{23}$$

$$MSE = \frac{1}{n}\sum_{i=1}^{n}(Target\ Values_i - Calculated\ Values_i)^2 \tag{24}$$

$$Penalty = \left\{ \begin{array}{ll} 0.3 \cdot \sum_{i=1}^{n} x_i\,, & if \sum_{i=1}^{n} x_i \notin [0.995, 1.01] \\ 0.1 \cdot MSE, & if Calculated\ Values_y \notin \left[OV_{low_y}, OV_{up_y}\right] \\ 0.1 \cdot MSE, & if Calculated\ Values_z \notin \left[OV_{low_z}, OV_{up_z}\right] \\ 0.1 \cdot \sum_{i=1}^{n} x_i\,, & if Calculated\ Values_n \notin \left[0, OV_{up_n}\right] \end{array} \right\} \tag{25}$$

The objective function is subject to:

$$\begin{bmatrix} x_1 \\ x_2 \\ \dots \\ x_{n-2} \end{bmatrix} \in \begin{bmatrix} 1 \cdot 10^{-6}, 0.3 \\ 1 \cdot 10^{-6}, 0.3 \\ \dots \\ 1 \cdot 10^{-6}, 0.3 \end{bmatrix} \bigwedge \begin{bmatrix} x_{n-1} \\ x_n \end{bmatrix} \in \begin{bmatrix} x_{low_{n-1}}, x_{up_{n-1}} \\ x_{low_n}, x_{up_n} \end{bmatrix} \tag{26}$$

Where $OV_{low}$ and $OV_{up}$ represent the boundaries that the calculated odor values should respect. The sub-indices y and z refer to the two scent families that are chosen for each circumstance. The sub-index n is relative to the alcoholic scent family. Similarly, the values of $x_{low}$ and $x_{up}$ represent the minimum and maximum molar fractions of solvents that are to be used in the new formulations. These should not differ too much from the original composition.

For the second perfume, the same optimization problem must be solved. However, since the target is different, the penalties included in the objective function must be tweaked so that the results are adequate. The first penalty, related to sum of the molar compositions, remains the same. The second and third penalties are altered to consider different and more suitable scent families and keep their odor values close to the target. The limits of the last constraint, relating to the alcoholic smell of the perfume are also tweaked in accordance with the alcoholic smell of the second fragrance. The values of the parameters used in this work are presented in Tables 2 and 3. The objective function and box constraints of the optimization problem remain the same.

Table 2 – Parameter values for both fragrances used as test cases in this work

| Sub-index | Parameter | First Fragrance | Second Fragrance |
|---|---|---|---|
| y | $OV_{low}$ | 0.11 | 0.23 |
| | $OV_{up}$ | 0.15 | 0.27 |
| z | $OV_{low}$ | 0.23 | 0.28 |
| | $OV_{up}$ | 0.27 | 0.32 |
| n-1 | $x_{low}$ | 0.1 | |
| | $x_{up}$ | 0.5 | |
| n | $OV_{up}$ | 0.18 | 0.2 |
| | $x_{low}$ | 0.55 | |

Table 3 – Explanation of sub-indices

| Sub-index | First fragrance | Second fragrance |
|---|---|---|
| y | Fruity | Sweet |
| z | Green | Coumarinic |
| n-1 | Water | |
| n | Ethanol | |
| | $x_{up}$ | 0.9 |

The next step is then to define the optimization algorithm that will solve the problems presented in Equations (22) to (26). In the realm of optimization, traditional gradient-based methods, which rely on calculating the gradient of the objective function, are often the first choice. Yet, when it comes to fragrance composition optimization, this path is fraught with challenges. The calculation of odor values, necessary for the objective function, lacks a readily available gradient, rendering conventional gradient-based optimization methods ineffective. As a result, an alternative approach becomes imperative. It is here that evolutionary optimization steps in as a viable solution.

Evolutionary optimization, a subfield of optimization inspired by the principles of natural selection and evolution, offers a compelling alternative to gradient-based methods. In this context, the Evolutionary.jl library (Art Wild, 2014) from Julia emerges as a valuable tool. Within this library, the Covariance Matrix Adaptation Evolution Strategy (CMA-ES) stands out as the preferred choice, having consistently demonstrated superior performance. CMA-ES, as an algorithm, specializes in addressing challenging optimization problems in continuous search spaces. It excels in handling complex scenarios characterized by non-convexity, ill-conditioning, multi-modality, rugged terrain, and noise. The adaptability and versatility of CMA-ES make it an ideal candidate for fine-tuning fragrance compositions in situations where traditional optimization methods fall short. It operates by iteratively evolving a population of candidate solutions, employing probabilistic models to guide the search towards the optimal composition. The synergy of evolutionary optimization principles and the computational prowess of CMA-ES enables the creation of fragrances that captivate the senses and fulfill the intended olfactory vision.

## 3. Results and discussion

The process of training the Gated Graph Neural Network (GGNN) involved the utilization of a database of fragrant molecules exhibiting diverse scent profiles. This training regimen spanned 150 epochs, during which a noteworthy observation was made. Epoch 115 emerged as the best candidate, as it recorded the lowest average train loss and validation loss. Consequently, this specific epoch was designated as the foundational point for all subsequent training phases in the transfer learning. The choice of this epoch as the starting point was not arbitrary but was guided by the monitors included in the GGNN, highlighting its efficiency in capturing and representing the essential fragrance characteristics within the model.

In the subsequent training phase, as previously outlined, the model was trained a second time on each of the scent families designated as target odors. To optimize the network's proficiency in generating molecules tailored to each specific scent family, a dedicated database containing only pertinent molecules was employed for each family. This phase entailed a training regimen lasting for 100 epochs, ensuring that the network matured in understanding the nuanced attributes of each fragrance profile. Notably, a pattern emerged, as it was consistently observed that the last epoch of each training cycle demonstrated the most promising results. This final epoch consistently exhibited the smallest losses, affirming the capacity of the model to discern and generate molecules that aligned with the target scent families.

Following the training phases, the GGNN was entrusted with the task of generating novel molecules, an endeavor that yielded valuable insights. Notably, the diversity and validity of the generated molecules were found to be influenced by the size of the specific databases employed during transfer learning. As anticipated, the use of more extensive databases in the training process translated into a greater capacity for the model to produce unique and valid molecules. This correlation between database size and output quality underscored the importance of database selection in the efficacy of the GGNN in generating molecules that met the desired fragrance profiles.

Given the necessary data for the calculation of odor values, the obtaining of vapor pressure for the generated molecules constituted a pivotal step in the subsequent analysis. This calculation was feasible solely for molecules with known thermodynamic parameters. As a result, the outcome of the calculation engendered a refined set of molecules, a cohort with known attributes by necessity, thus allowing for further analysis to be made. Table 2 gives an overview of the refinement results.

Table 5 – Information on the generated molecules.

| Fragrance notes | Number of unique and valid molecules | Number of molecules with known vapor pressure |
|---|---|---|
| Citrus | 51 | 15 |
| Sweet | 101 | 16 |
| Fruity | 107 | 28 |
| Coumarinic | 131 | 27 |
| Sharp | 15 | 6 |
| Floral | 32 | 20 |
| Green | 8 | 4 |
| Fatty | 63 | 35 |
| Spicy | 82 | 15 |
| Musk | 138 | 27 |
| Wood | 18 | 7 |
| Lemon | 36 | 8 |
| Rose | 92 | 19 |

From this refined list, one molecule was chosen from each family, with the closest vapor pressure to the original molecule, and their scent was consulted in The Good Scents Company to confirm that the target scent we want is achieved. To be in conformity with the target scent, a main fragrance note, and a secondary fragrance note for each molecule was included. Two important observations were made. While the descriptor used in the GGNN generally meant that the molecule being generated would contain that descriptor as a fragrance note, it was not necessarily the main fragrance note of that molecule. The first molecule in Table 3, corresponding to the "citrus" descriptor serves as an example: the descriptor is shown as the secondary fragrance note. It was also observed that some of the selected molecules already contained other target fragrance notes in their fragrance profile. For that reason, the descriptors "rose", "fatty" and "fruity" were not used for the formulation of the new mixture, while the descriptor "floral" was used twice, as the molecules generated for that word were quite varied.

Table 6 – Information on the selected molecules.

| Descriptor used in GGNN | SMILES | MW (g/mol) | Main Fragrance Note | Secondary Fragrance Note | $P_{sat}$ (Pa) |
|---|---|---|---|---|---|
| Citrus | CCCCCCCCCC=O | 156.26 | Sweet | Citrus | 10.431 |
| Sweet | CCCCCCCC(=O)OCCC(C)C | 214.34 | Sweet | Fruity | 0.946 |
| Coumarin | C1CC(=O)OC2=CC=CC=C21 | 148.16 | Coumarin | Sweet | 1.260 |
| Spicy | CCCCCCC(C)O | 130.23 | Spicy | Green | 8.729 |
| Sharp | CCOC(=O)C(C)O | 118.13 | Sharp | Fruity | 42.135 |
| Floral | CCCCC(CC)CO | 130.23 | Citrus | Floral | 21.646 |
| Green | CCC=CCCO | 100.16 | Green | Green | 130.077 |
| Lemon | CCCCCC=CC=CC(=O)OC | 182.26 | Fruity | Lemon | 2.565 |
| Musk | CCCCCCCCCCCCCC=O | 212.37 | Fatty | Musk | 0.752 |
| Floral | CCCCCCCCCCO | 158.28 | Floral | Rose | 2.456 |
| Wood | CCCCCCCO | 116.20 | Green | Wood | 24.986 |
| Ethanol | CCO | 46 | Alcoholic | Alcoholic | 7270 |
| Water | O | 18 | Odorless | Odorless | 3170 |

Upon obtaining all the required information, the optimization algorithm was run. Given the inherent stochasticity of the Covariance Matrix Adaptation Evolution Strategy (CMA-ES), several iterations were necessary to attain results that met the criteria for both fragrances. The culmination of this optimization is briefly summarized in Tables 4 and 5, providing a comprehensive insight into the refined fragrance compositions achieved through the iterative CMA-ES process.

Table 7 – Optimized compositions of new mixtures.

| SMILES | Optimized Molar Composition for First Fragrance | Optimized Molar Composition for Second Fragrance |
|---|---|---|
| CCCCCCCCCC=O | $7.48 \times 10^{-4}$ | $2.20 \times 10^{-3}$ |
| CCCCCCCC(=O)OCCC(C)C | $7.66 \times 10^{-4}$ | $9.17 \times 10^{-3}$ |
| C1CC(=O)OC2=CC=CC=C21 | $2.85 \times 10^{-2}$ | $1.74 \times 10^{-1}$ |
| CCCCCCC(C)O | $8.70 \times 10^{-4}$ | $4.02 \times 10^{-3}$ |
| CCOC(=O)C(C)O | $5.03 \times 10^{-2}$ | $2.20 \times 10^{-3}$ |
| CCCCC(CC)CO | $4.47 \times 10^{-2}$ | $3.21 \times 10^{-3}$ |
| CCC=CCCO | $9.48 \times 10^{-2}$ | $6.60 \times 10^{-3}$ |
| CCCCCC=CC=CC(=O)OC | $3.51 \times 10^{-2}$ | $2.20 \times 10^{-3}$ |
| CCCCCCCCCCCCCC=O | $9.57 \times 10^{-2}$ | $2.85 \times 10^{-3}$ |
| CCCCCCCCCCO | $1.04 \times 10^{-4}$ | $6.75 \times 10^{-3}$ |
| CCCCCCCO | $2.47 \times 10^{-4}$ | $2.20 \times 10^{-3}$ |
| CCO | $1.00 \times 10^{-1}$ | $1.00 \times 10^{-1}$ |
| O | $5.53 \times 10^{-1}$ | $6.85 \times 10^{-1}$ |

Table 8 – Calculated and target odor values for new mixtures.

| Fragrance Note | Target Odor Values for First Fragrance | Odor Values for Optimized Composition | | Target Odor Values for Second Fragrance | Odor Values for Optimized Composition |
|---|---|---|---|---|---|
| Citrus | $5.79 \times 10^{-2}$ | $7.57 \times 10^{-2}$ | | $3.98 \times 10^{-3}$ | $2.97 \times 10^{-2}$ |
| Sweet | $8.87 \times 10^{-2}$ | $1.20 \times 10^{-2}$ | | $2.51 \times 10^{-1}$ | $2.06 \times 10^{-1}$ |
| Fruity | $1.33 \times 10^{-1}$ | $1.10 \times 10^{-1}$ | | $2.06 \times 10^{-2}$ | $1.86 \times 10^{-2}$ |
| Coumarinic | $1.41 \times 10^{-1}$ | $2.66 \times 10^{-2}$ | | $3.03 \times 10^{-1}$ | $4.23 \times 10^{-1}$ |
| Sharp | $7.75 \times 10^{-2}$ | $2.20 \times 10^{-1}$ | | $1.01 \times 10^{-2}$ | $1.54 \times 10^{-2}$ |
| Floral | $2.49 \times 10^{-2}$ | $3.24 \times 10^{-2}$ | | $3.91 \times 10^{-2}$ | $1.98 \times 10^{-2}$ |
| Green | $2.51 \times 10^{-1}$ | $3.21 \times 10^{-1}$ | | $4.41 \times 10^{-2}$ | $4.09 \times 10^{-2}$ |
| Fatty | $8.34 \times 10^{-4}$ | $1.04 \times 10^{-2}$ | | $5.95 \times 10^{-4}$ | $2.27 \times 10^{-3}$ |
| Spicy | $7.73 \times 10^{-3}$ | $5.82 \times 10^{-4}$ | | $3.89 \times 10^{-2}$ | $9.43 \times 10^{-3}$ |
| Musk | $3.57 \times 10^{-4}$ | $4.47 \times 10^{-3}$ | | $2.55 \times 10^{-4}$ | $9.73 \times 10^{-4}$ |
| Woody | $8.70 \times 10^{-3}$ | $1.22 \times 10^{-4}$ | | $1.11 \times 10^{-2}$ | $3.26 \times 10^{-3}$ |
| Lemon | $3.32 \times 10^{-2}$ | $6.62 \times 10^{-3}$ | | $4.34 \times 10^{-3}$ | $1.43 \times 10^{-3}$ |
| Rose | $3.52 \times 10^{-5}$ | $1.01 \times 10^{-5}$ | | $1.13 \times 10^{-4}$ | $3.36 \times 10^{-3}$ |
| Alcoholic | $1.76 \times 10^{-1}$ | $1.80 \times 10^{-1}$ | | $1.82 \times 10^{-1}$ | $2.26 \times 10^{-1}$ |

For a more in-depth understanding of the transformation wrought by this optimization, a selection of six distinct fragrance notes was chosen as illustrative exemplars. The visuals, encapsulated in Figures 4 and 5, depict these olfactory nuances. They represent visually the quality of the reproduction executed in this work and allows for a comparison of the target fragrances presented in the literature.

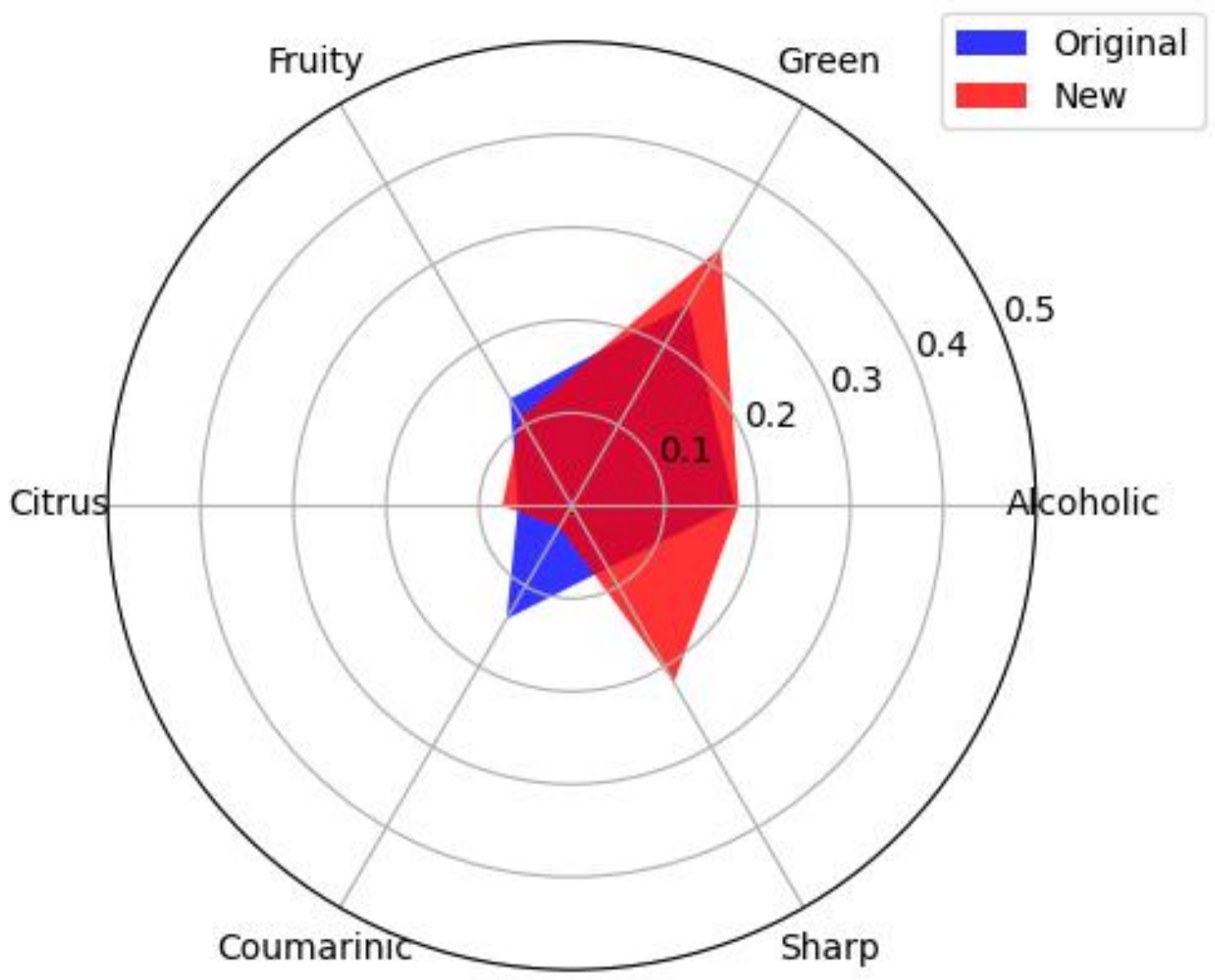

Figure 4 – Comparison of six selected fragrance notes for fragrance 1.

An empirical analysis of fragrance 1 reveals a noteworthy resemblance between most odor values and their corresponding target values. This observation underscores an alignment between the newly formulated fragrance mixture and the original perfume, with a few differences. However, there are some notable differences, the clearest being the distinction between the odor values of the coumarinic and sharp families. Since the newly generated molecules have different fragrance profiles and properties than the original molecules, some variations between the new mixture and the target are unavoidable.

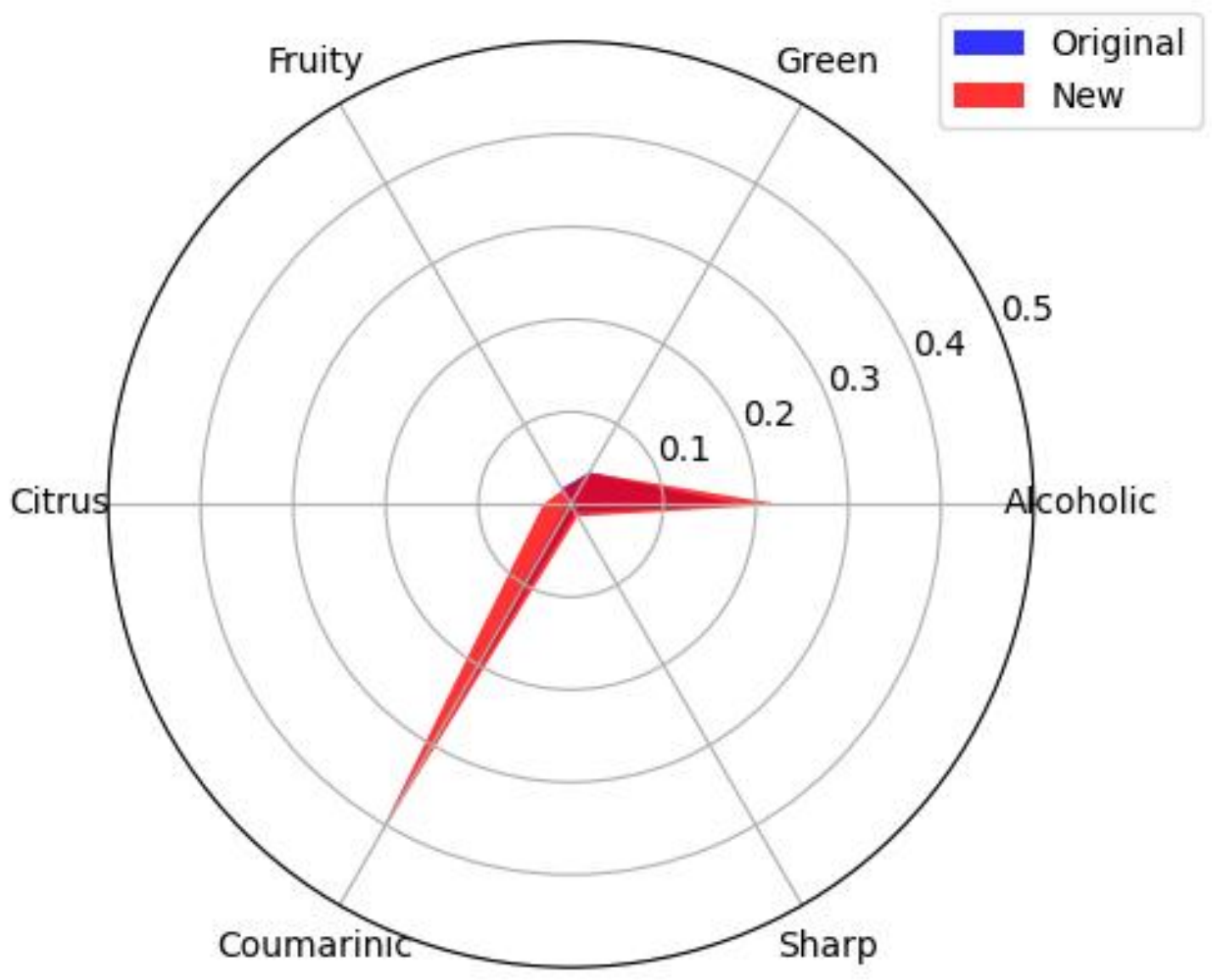

Figure 5 – Comparison of six selected fragrance notes for fragrance 2.

Regarding the second fragrance, the methodology here proposed can recreate the smell with high accuracy, as made evident by Figure 5.

The mean squared error for the first fragrance is computed to be 3.28 $x10^{-3}$, signifying the degree of deviation between the target and calculated odor values, as presented in equation (24). Similarly, for the second fragrance, the mean squared error is calculated to be 1.45 $x10^{-3}$. These quantifiable metrics provide a numerical lens through which the efficacy of the optimization process can be gauged, offering a quantitative measure of the disparities between the original and optimized fragrance compositions.

The inherent diversity in fragrance profiles of newly generated molecules inevitably introduces small distinctions between the original and optimized fragrances for the first fragrance. In this context, natural variations are discerned, particularly in the intensities of Coumarinic and Sharp notes. Notably, the first original fragrance exhibits a higher intensity of Coumarinic notes compared to Sharp notes, a dynamic that is inverted in the optimized perfume. While other less intense notes also exhibit disparities between the two perfumes, their relatively smaller odor values imply that the perceptible difference in fragrance may not be as pronounced.

A comprehensive examination of the results for the second fragrance indicates a precise reproduction. A substantial concordance is observed in the odor values of most fragrance notes. Additionally, the visual representation in Figure 5 corroborates this finding, revealing a consistent intensity relationship between Coumarinic and Sharp notes in both the original and optimized fragrances. Although the initial reproduction wasn't flawless, it provides perfumers with a valuable starting point, significantly easing the specialist's workload. In contrast, the second fragrance is highly satisfactory as is.

Given that the principles here demonstrated are validated, a natural next step could involve augmenting the model's training with larger databases and generating a greater number of molecules. This approach could facilitate the selection of molecules with fragrance profiles more closely aligned with the originals, thereby refining the optimization process.

## 4. Conclusion

The traditional way of creating new scents by trial and error is still very labor intensive and time consuming. Although there have been innovations in the area in the last years, there is still a lot of room for new solutions that improve the efficiency of the creation process. The method presented in this work is intended to be used as a tool in the manufacturing of new perfumes. In this work, two target fragrances were quantified by calculating the odor value of their constituents and then calculating the odor value for the scent families present in the perfume. Then, fragrant molecules were generated to substitute the ingredients that made up the fragrance notes in the original perfume. The best candidates were selected form the set of generated molecules by estimating their vapor pressure and consulting their fragrance profile. Finally, a new fragrance was suggested by optimizing the composition of a new mixture with the newly generated fragrance molecules.

The analysis of fragrance 1 revealed a predominantly close alignment between the calculated odor values and the target values, signifying a favorable correspondence between the new fragrance mixture and the original perfume. Due to intrinsic differences in the fragrance profiles of the generated and the original molecules, a few distinctions in aroma were observed. Notably, a significant difference is observed in the intensity relationship between Coumarinic and Sharp notes, with the original fragrance showcasing a higher intensity of Coumarinic notes, whereas this dynamic is reversed in the optimized perfume. In contrast, the second fragrance exhibits more consistent results, with most fragrance notes displaying similar odor values, and a visual analysis corroborates the absence of intensity switches between Coumarinic and Sharp notes. The mean squared error for the first fragrance was calculated as $3.28 \times 10^{-3}$, and for the second fragrance, it was $1.45 \times 10^{-3}$.

While the method displays promise, there is room for improvement. Augmenting the model's training with larger databases and generating a greater number of molecules may enhance the ability to select compounds with fragrance profiles more closely aligned with the originals, resulting in a better match between the new and the original mixtures.

## 5. References

Almeida, R. N., Costa, P., Pereira, J., Cassel, E., & Rodrigues, A. E. (2019). Evaporation and Permeation of Fragrance Applied to the Skin. *Industrial & Engineering Chemistry Research*, *58*(22), 9644–9650. https://doi.org/10.1021/acs.iecr.9b01004

Ambrose, D., & Walton, J. (1989). Vapour pressures up to their critical temperatures of normal alkanes and 1-alkanols. *Pure and Applied Chemistry*, *61*(8), 1395–1403. https://doi.org/10.1351/pac198961081395

Appell, L. (1969). *Physical foundations in perfumery* (84th ed., pp. 45–50). Am. Perfum. Cosmet.

Art Wild. (2014). *Evolutionary.jl*. https://wildart.github.io/Evolutionary.jl/stable/

Arús-Pous, J., Johansson, S. V., Prykhodko, O., Bjerrum, E. J., Tyrchan, C., Reymond, J.-L., Chen, H., & Engkvist, O. (2019). Randomized SMILES strings improve the quality of molecular generative models. *Journal of Cheminformatics*, *11*(1), 71. https://doi.org/10.1186/s13321-019-0393-0

Bell, C. and C. (2023). *Thermo: Chemical properties component of Chemical Engineering Design Library (ChEDL)*. https://github.com/CalebBell/thermo.

Bembibre, C., & Strlič, M. (2017). Smell of heritage: a framework for the identification, analysis and archival of historic odours. *Heritage Science*, *5*(1), 2. https://doi.org/10.1186/s40494-016-0114-1

Bembibre, C., & Strlič, M. (2022). From Smelly Buildings to the Scented Past: An Overview of Olfactory Heritage. *Frontiers in Psychology*, *12*. https://doi.org/10.3389/fpsyg.2021.718287

Bushdid, C., Magnasco, M. O., Vosshall, L. B., & Keller, A. (2014). Humans can discriminate more than 1 trillion olfactory stimuli. *Science*, *343*(6177), 1370–1372. https://doi.org/10.1126/science.1249168

Carles, J. (1961). *A Method of Creation & Perfumery*.

Curran, K., Underhill, M., Grau-Bové, J., Fearn, T., Gibson, L. T., & Strlič, M. (2018). Classifying Degraded Modern Polymeric Museum Artefacts by Their Smell. *Angewandte Chemie International Edition*, *57*(25), 7336–7340. https://doi.org/10.1002/anie.201712278

Debnath, T., & Nakamoto, T. (2022). Predicting individual perceptual scent impression from imbalanced dataset using mass spectrum of odorant molecules. *Scientific Reports*, *12*(1). https://doi.org/10.1038/s41598-022-07802-3

Devos, M., Patte, F., Rouault, J., Laffort, P., & Gemert, L. J. (1990). *Standardized Human Olfactory Thresholds*. IRL Press.

Fortune Business Insight. (2022, May). *Flavors and Fragrances Market Size, Share & COVID-19 Impact Analysis*.

Gerkin, R. C. (2021). Parsing Sage and Rosemary in Time: The Machine Learning Race to Crack Olfactory Perception. In *Chemical Senses* (Vol. 46). Oxford University Press. https://doi.org/10.1093/chemse/bjab020

Grau-Bové, J., & Strlič, M. (2013). Fine particulate matter in indoor cultural heritage: a literature review. *Heritage Science*, *1*(1), 8. https://doi.org/10.1186/2050-7445-1-8

Heng, Y. P., Lee, H. Y., Chong, J. W., Tan, R. R., Aviso, K. B., & Chemmangattuvalappil, N. G. (2022). Incorporating Machine Learning in Computer-Aided Molecular Design for Fragrance Molecules. *Processes*, *10*(9), 1767. https://doi.org/10.3390/pr10091767

Herz, R. S., Schankler, C., & Beland, S. (2004). Olfaction, Emotion and Associative Learning: Effects on Motivated Behavior. *Motivation and Emotion*, *28*(4), 363–383. https://doi.org/10.1007/s11031-004-2389-x

Kingma, D. P., & Ba, J. (2014). *Adam: A Method for Stochastic Optimization*.

Kullback, S., & Leibler, R. A. (1951). On Information and Sufficiency. *The Annals of Mathematical Statistics*, *22*(1), 79–86. http://www.jstor.org/stable/2236703

Lee, B. I., & Kesler, M. G. (1975). A generalized thermodynamic correlation based on three-parameter corresponding states. *AIChE Journal*, *21*(3), 510–527. https://doi.org/10.1002/aic.690210313

Leffingwell & Associates. (2018). *Flavor & Fragrance Industry - Top 10*. Flavor & Fragrance Industry - Top 10

Li, Y., Tarlow, D., Brockschmidt, M., & Zemel, R. (2015). *Gated Graph Sequence Neural Networks*.

Ludmir, C. (2023, July 31). *Fragrance Players Enjoy Double-Digit Growth, Driven By Demand For Premium Labels*. Forbes. https://www.forbes.com/sites/claraludmir/2023/07/31/fragrance-players-enjoy-double-digit-growth-driven-by-demand-for-premium-labels/

Mata, V. G., Gomes, P. B., & Rodrigues, A. E. (2005). Engineering perfumes. *AIChE Journal*, *51*(10), 2834–2852. https://doi.org/10.1002/aic.10530

Mullol, J., Alobid, I., Mariño-Sánchez, F., Izquierdo-Domínguez, A., Marin, C., Klimek, L., Wang, D.-Y., & Liu, Z. (2020). The Loss of Smell and Taste in the COVID-19 Outbreak: a Tale of Many Countries. *Current Allergy and Asthma Reports*, *20*(10), 61. https://doi.org/10.1007/s11882-020-00961-1

Nozaki, Y., & Nakamoto, T. (2018). Predictive modeling for odor character of a chemical using machine learning combined with natural language processing. *PLoS ONE*, *13*(6). https://doi.org/10.1371/journal.pone.0198475

Ooi, Y. J., Aung, K. N. G., Chong, J. W., Tan, R. R., Aviso, K. B., & Chemmangattuvalappil, N. G. (2022). Design of fragrance molecules using computer-aided molecular design with machine learning. *Computers & Chemical Engineering*, *157*, 107585. https://doi.org/10.1016/j.compchemeng.2021.107585

Poling, B., Prausnitz, J., & O'Connell, J. (2004). *The Properties of Gases and Liquids* (5th ed.). McGraw-Hill.

Poucher, W. A. (1955). A classification of odors and its uses. *J. Soc. Cosmet. Chem.*, 81–95.

Prosen, H., Janeš, L., Strlič, M., Rusjan, D., & Kočar, D. (2007). Analysis of Free and Bound Aroma Compounds in Grape Berries Using Headspace Solid-Phase Microextraction with GC-MS and a Preliminary Study of Solid-Phase Extraction with LC-MS. *Acta Chimica Slovenica*.

Queiroz, L. P., Rebello, C. M., Costa, E. A., Santana, V. V., Rodrigues, B. C. L., Rodrigues, A. E., Ribeiro, A. M., & Nogueira, I. B. R. (2023a). A Reinforcement Learning Framework to Discover Natural Flavor Molecules. *Foods*, *12*(6), 1147. https://doi.org/10.3390/foods12061147

Queiroz, L. P., Rebello, C. M., Costa, E. A., Santana, V. V., Rodrigues, B. C. L., Rodrigues, A. E., Ribeiro, A. M., & Nogueira, I. B. R. (2023b). Generating Flavor Molecules Using Scientific Machine Learning. *ACS Omega*, *8*(12), 10875–10887. https://doi.org/10.1021/acsomega.2c07176

Queiroz, L. P., Rebello, C. M., Costa, E. A., Santana, V. V., Rodrigues, B. C. L., Rodrigues, A. E., Ribeiro, A. M., & Nogueira, I. B. R. (2023c). Transfer Learning Approach to Develop Natural Molecules with Specific Flavor Requirements. *Industrial & Engineering Chemistry Research*, *62*(23), 9062–9076. https://doi.org/10.1021/acs.iecr.3c00722

Radhakrishnapany, K. T., Wong, C. Y., Tan, F. K., Chong, J. W., Tan, R. R., Aviso, K. B., Janairo, J. I. B., & Chemmangattuvalappil, N. G. (2020). Design of fragrant molecules through the incorporation of rough sets into computer-aided molecular design.

*Molecular Systems Design & Engineering*, *5*(8), 1391–1416. https://doi.org/10.1039/D0ME00067A

*RDKit: Open-source cheminformatics*. (n.d.). https://www.rdkit.org

Rodrigues, A. E., Nogueira, I., & Faria, R. P. V. (2021). Perfume and Flavor Engineering: A Chemical Engineering Perspective. *Molecules*, *26*(11), 3095. https://doi.org/10.3390/molecules26113095

Rodrigues, B. C. L., Santana, V. V, Queiroz, L. P., Rebello, C. M., & Nogueira, I. B. R. (2023). *Scents of AI: Harnessing Graph Neural Networks to Craft Fragrances Based on Consumer Feedback*. https://doi.org/10.20944/preprints202310.0247.v1

Rodríguez, O., Teixeira, M. A., & Rodrigues, A. E. (2011). Prediction of odour detection thresholds using partition coefficients. *Flavour and Fragrance Journal*, *26*(6), 421–428. https://doi.org/10.1002/ffj.2076

Royal Society of Chemistry. (n.d.). *ChemSpider*.

Saini, K., & Ramanathan, V. (2022). *A Review of Machine Learning Approaches to Predicting Molecular Odor in the Context of Multi-Label Classication*. https://doi.org/10.21203/rs.3.rs-1492792/v1

Sanchez-Lengeling, B., Wei, J. N., Lee, B. K., Gerkin, R. C., Aspuru-Guzik, A., & Wiltschko, A. B. (2019). *Machine Learning for Scent: Learning Generalizable Perceptual Representations of Small Molecules*.

Santana, V. V., Martins, M. A. F., Loureiro, J. M., Ribeiro, A. M., Rodrigues, A. E., & Nogueira, I. B. R. (2021). Optimal fragrances formulation using a deep learning neural network architecture: A novel systematic approach. *Computers & Chemical Engineering*, *150*, 107344. https://doi.org/10.1016/j.compchemeng.2021.107344

Sharma, A., Kumar, R., Ranjta, S., & Varadwaj, P. K. (2021). SMILES to Smell: Decoding the Structure–Odor Relationship of Chemical Compounds Using the Deep Neural Network Approach. *Journal of Chemical Information and Modeling*, *61*(2), 676–688. https://doi.org/10.1021/acs.jcim.0c01288

Synnott, A. (1991). A sociology of smell. *Canadian Review of Sociology/Revue Canadienne de Sociologie*, *28*(4), 437–459. https://doi.org/10.1111/j.1755-618X.1991.tb00164.x

Teixeira, M. A., Barrault, L., Rodríguez, O., Carvalho, C. C., & Rodrigues, A. E. (2014). Perfumery radar 2.0: A step toward fragrance design and classification. *Industrial and Engineering Chemistry Research*, *53*(21), 8890–8912. https://doi.org/10.1021/ie403968w

Teixeira, M. A., Rodríguez, O., Mata, V. G., & Rodrigues, A. E. (2009). The diffusion of perfume mixtures and the odor performance. *Chemical Engineering Science*, *64*(11), 2570–2589. https://doi.org/10.1016/j.ces.2009.01.064

Teixeira, M. A., Rodríguez, O., & Rodrigues, A. E. (2010). Perfumery radar: A predictive tool for perfume family classification. *Industrial and Engineering Chemistry Research*, *49*(22), 11764–11777. https://doi.org/10.1021/ie101161v

Teixeira, M., Rodríguez, O., Gomes, P., Mata, V., & Rodrigues, A. (2013). *Perfume Engineering. Design, Performance & Classification.* Butterworth-Heinemann.

*The Good Scents Company*. (2021). http://www.thegoodscentscompany.com/

van Gemert, L. J. (2003). *Compilations of odour threshold values in air, water and other media*. Oliemans Punter & Partners BV.

Wakayama, H., Sakasai, M., Yoshikawa, K., & Inoue, M. (2019). Method for Predicting Odor Intensity of Perfumery Raw Materials Using Dose–Response Curve Database. *Industrial & Engineering Chemistry Research*, *58*(32), 15036–15044. https://doi.org/10.1021/acs.iecr.9b01225

Walker, P. J., Yew, H.-W., & Riedemann, A. (2022). *Clapeyron.jl: An extensible, open-source fluid-thermodynamics toolkit*.

Weininger, D., Weininger, A., & Weininger, J. L. (1989). SMILES. 2. Algorithm for generation of unique SMILES notation. *Journal of Chemical Information and Computer Sciences*, *29*(2), 97–101. https://doi.org/10.1021/ci00062a008

Xue, D., Gong, Y., Yang, Z., Chuai, G., Qu, S., Shen, A., Yu, J., & Liu, Q. (2019). Advances and challenges in deep generative models for de novo molecule generation. *WIREs Computational Molecular Science*, *9*(3). https://doi.org/10.1002/wcms.1395

Zhang, L., Mao, H., Liu, L., Du, J., & Gani, R. (2018). A machine learning based computer-aided molecular design/screening methodology for fragrance molecules. *Computers & Chemical Engineering*, *115*, 295–308. https://doi.org/10.1016/j.compchemeng.2018.04.018

Zhou, Z., Kearnes, S., Li, L., Zare, R. N., & Riley, P. (2019). Optimization of Molecules via Deep Reinforcement Learning. *Scientific Reports*, *9*(1), 10752. https://doi.org/10.1038/s41598-019-47148-x